# Thermal Spin Valves

*Brian Sales, Rongying Jin, and David Mandrus*

Condensed Matter Sciences Division, Oak Ridge National Laboratory
Oak Ridge, TN 37831-6056

**ABSTRACT**

The ability of an insulating solid to conduct heat is rarely effected by the application of a magnetic field. We have found, however, that the low temperature heat conduction of some solids increases by more than a factor of two with the application of a modest magnetic field. The effect occurs in low-dimensional magnetically ordered materials when a small gap, $\Delta$, in the acoustic magnon (spin-wave) spectra is closed using a magnetic field $H > \Delta/g\mu_B$. Since all magnetically ordered materials must have a gap in the magnon spectra for magnons with $k \approx 0$, this may be a very general effect. Extra heat is carried through the solid only when the magnetic field exceeds the critical value ($H > \Delta/g\mu_B$). At this critical field the tiny atomic magnets in the solid abruptly change the direction they point which results in more heat flowing through the material. The magnetic field thus acts as a heat switch. We have observed this effect in three quite different magnetically ordered materials: $K_2V_3O_8$, $Nd_2CuO_4$ and $Pr_2CuO_4$. Several possible explanations for these effects will be discussed.

**INTRODUCTION**

Thermal conductivity measurements have been used as a probe of various magnetic systems on and off for the past 40 years[1-3]. Recently there has been increased interest in low-dimensional magnets and the relationship between magnetism and superconductivity. As a result, in insulating magnetic systems there has been renewed interest in using thermal conductivity measurements to probe excitations in a fashion analogous to the use of electrical conductance measurements to probe metallic systems [4,5]. The insulating spin-ladder compound $Ca_9La_5Cu_{24}O_{41}$, for example, provides perhaps the clearest evidence of a huge magnon ("spinon") contribution to the heat current. Along the spin-ladder direction the magnetic portion of the thermal conductivity at room temperature is comparable to the thermal conductivity of metallic platinum (≈ 140 W/m-K)[6]! For several unconventional superconductors, magneto-thermal conductivity measurements also have been used to probe the symmetry of the superconducting gap [7] and the possible breakdown of the Wiedemann-Franz Law and Fermi Liquid Theory [8].

In the present article we summarize some of our recent results [9, 10] on the magnetic field dependence of the low temperature thermal conductivity of three quasi-2d antiferromagnets, $K_2V_3O_8$, $Nd_2CuO_4$ and $Pr_2CuO_4$.

**EXPERIMENTAL ASPECTS**

Single crystals of $K_2V_3O_8$ (typical dimensions 5 x 5 x 0.5 mm$^3$) are grown in a Pt crucible sealed in a silica ampoule using a $KVO_3$ flux. Crystals of $Nd_2CuO_4$ and $Pr_2CuO_4$ are grown using both the travelling solvent float zone method and the flux method with excess CuO as the flux or solvent (typical dimensions 5 x 2 x 1 mm$^3$). Four point thermal conductivity measurements are made with a Physical Property Measurement System (PPMS) from Quantum Design. Four 0.020" diameter copper wires are attached to each sample using silver epoxy (H20E from EPOTEK). The cold finger, sample heater and the Cernox thermometry "shoes" are clamped to the copper wires. The type of Cernox thermometers we use are relatively insensitive to external magnetic fields. We estimate that at 2 K and 9T the error in the measured thermal conductivity is less than 4% (due to the thermometry) and much less at higher temperatures or lower magnetic fields. We find that for isothermal measurements as a function of field the "timed mode" of the PPMS thermal conductivity option gives much better data than the fully automated mode, particularly at the lowest temperatures.

**RESULTS**

$K_2V_3O_8$ crystallizes in a tetragonal unit cell with space group P4*bm* and *a* = 8.87 Å and *c* = 5.215 Å. The magnetic S=1/2 V$^{+4}$ atoms form a simple 2D square lattice (see Fig 1). The powder magnetic susceptibility data above 4 K are well described by a 2D Heisenberg model with coupling constant J =12.6 K [11]. Lumsden et al. [12] used a combination of neutron scattering and magnetic susceptibility data from single crystals to show that $K_2V_3O_8$ orders magnetically at about 4 K and exhibits weak ferromagnetism (very slightly canted antiferromagnetic order) and field-induced spin reorientations at fields of $H_c$ = 0.65 T for H // *a*, and $H_c$ = 0.95 T for H // *c*. At $H_c$, the small gap Δ in the magnon spectra at *k* = 0 is closed and a new magnetic ground state is formed. The rotation of the V$^{+4}$ spins from along the *c* axis, into the plane for H > $H_c$ results in an additional channel for heat conduction (Fig 1).

$Nd_2CuO_4$ crystallizes in a tetragonal unit cell I4/mmm with *a* = 3.94 Å and *c* = 12.17 Å. Above room temperature the magnetic susceptibility of $Nd_2CuO_4$ can be described by a 2D Heisenberg model with a coupling constant J ≈ 1500 K between Cu spins in the same layer. The Cu sublattice orders antiferromagnetically at $T_N$ ≈

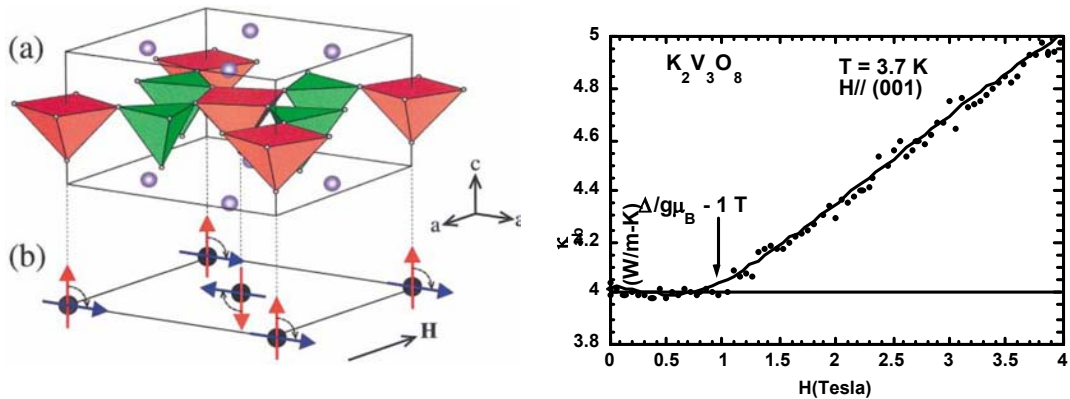

Figure 1. (Left a) Crystal structure of $K_2V_3O_3$ consists of magnetic pyramids of $V^{+4}$-$O_5$ and non magnetic $V^{+5}$-$O_4$ tetrahedra. (Left b) Projection of the magnetic $V^{+4}$ positions into the basal plane showing the locations of the magnetic moments. In zero magnetic field the spins point along the $c$ axis. A field of 1 T along the $c$ axis or 0.65 T along the $a$ axis causes the spins to flop into the basal plane. (Right) Effect of spin flop on thermal conductivity.

280 K. On further cooling, the large exchange coupling between the Cu and Nd polarizes the Nd spins and induces an ordered moment. Below about 30 K the ordered arrangement of the Cu and Nd spins is non-collinear (Fig 2). The induced moment on the Nd is about 1.5 $\mu_B$ at 2 K and 0.05 $\mu_B$ at 50 K. At 5 K, the application of a field $H_c \approx 5$ T along the [100] direction results in a spin-flop transition in which the Nd and Cu spins along the [100] direction rotate in plane by 90°. This results in a collinear magnetic structure in which all of the moments are perpendicular to H. The magnetic excitation spectra of $Nd_2CuO_4$ [13, 14] is much more complicated than that of $K_2V_3O_8$ which has only one type of magnetic ion per unit cell. Nevertheless, the transition in $Nd_2CuO_4$ from a non-collinear to a collinear magnetic structure implies that the gap at $k = 0$ in one of the magnon branches goes to 0 at $H_c \approx 5$ T. As was the case for $K_2V_3O_8$, for $H > H_c \approx 5$ T, the in-plane thermal conductivity of $Nd_2CuO_4$ exhibits an additional channel for heat conduction (Fig 2). We tried to measure the thermal conductivity of $Nd_2CuO_4$ at low temperatures with H//c, since with a magnetic field in this direction no spin flop is observed using neutron scattering. The large anisotropy in the magnetic susceptibility for H//a or H//c, however, creates a large torque on the sample when H is applied along $c$. This torque is sufficient to bend our Cu support piece making measurements difficult.

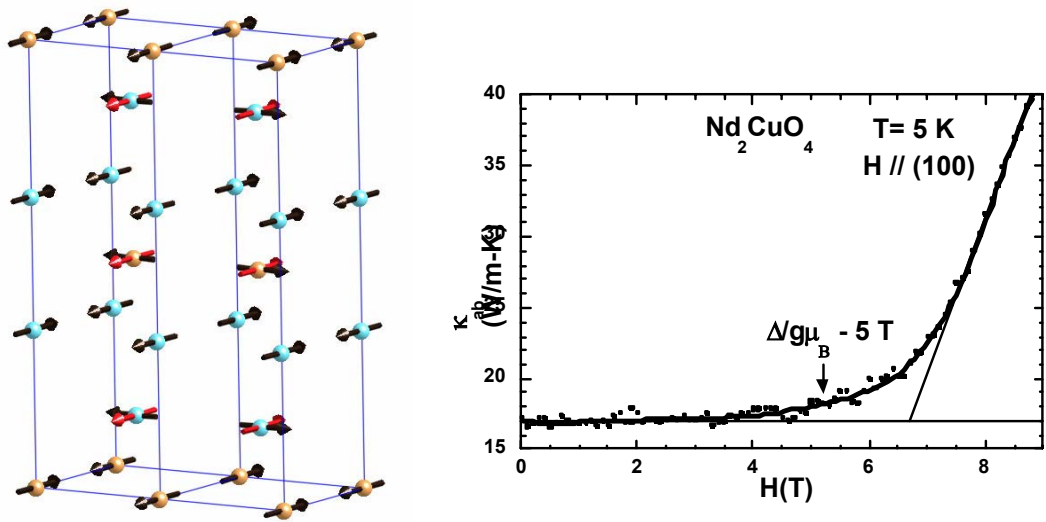

Figure 2. (Left) Magnetic structure of $Nd_2CuO_4$. Three planes of Cu spins (tan) are shown at the top middle and bottom of the drawing. Four layers of Nd spins (light blue) also are shown. In zero magnetic field the Cu spins in adjacent layers are rotated by 90° as are the Nd spins. A transition to a collinear structure occurs for $H_c \approx 5$ T. (Right) The effect of this spin-flop transition on the in-plane thermal conductivity.

$Pr_2CuO_4$ crystallizes in the same structure as $Nd_2CuO_4$ with $a = 3.95$ Å and $c = 12.18$. The magnetic interactions are also similar with in-plane $J_{Cu-Cu} \approx 1500$ K and $T_N \approx 250$ K. At low temperatures the Cu (0.4 $\mu_B$) and Pr (0.08 $\mu_B$) moments order in a non-collinear magnetic structure that is similar (but not identical) to the structure adopted by $Nd_2CuO_4$ [13]. For $Pr_2CuO_4$, neutron data shows a similar non-collinear to collinear spin-flop transition [15] for H// [110] direction. In $Pr_2CuO_4$, $H_c$ increases rapidly with decreasing temperature and reaches a value of 4.5 T at T $\approx$ 2 K [15]. In-plane thermal conductivity measurements with H//[100] show an increase in κ starting at a much lower magnetic field of H $\approx$ 0.5 T. Although the field of the spin-flop transition may be somewhat sample dependent, it seems more likely that there are *other* excitations involving the Pr ions that are responsible for the effects shown in Fig. 3.

The lowest crystal-field-level of the Pr ion is a non-magnetic singlet with the first excited level, a magnetic doublet, at about 200 K. At 2 K, however, neutron data show an induced moment of about 0.08 $\mu_B$ per Pr. Such an induced moment is only possible with an admixture of higher J (angular momentum) mutiplets and/or if the Cu exchange field mixes one of the Pr excited states into the singlet ground state. The neutron data [15] also indicates relatively strong Pr-Pr exchange and dispersion *both* in-plane and along the $c$ direction through *magnetic excitons*. Magnetic excitons are the elementary excitations in singlet ground state systems and in Pr metal are known to strongly couple to phonons in an external magnetic field [15]. To test some of these ideas, we measured the in-plane thermal conductivity of $Pr_2CuO_4$ with the magnetic field applied along the *c* axis (Fig. 4). The large increase in κ indicates that the spin-flop transition is not the only mechanism impacting the thermal conductivity of this material in a magnetic field. We note that since this effect is most likely associated with the Pr sublattice, a similar effect probably occurs in the Ce-doped superconductor ($Pr_{1.85}Ce_{0.15}CuO_4$). A large change in the non-electronic portion of the thermal conductivity in a magnetic field clearly brings into question the conclusions of Hill et al. [8] concerning the breakdown of the Wiedemann-Franz law in $Pr_{1.85}Ce_{0.15}CuO_4$.

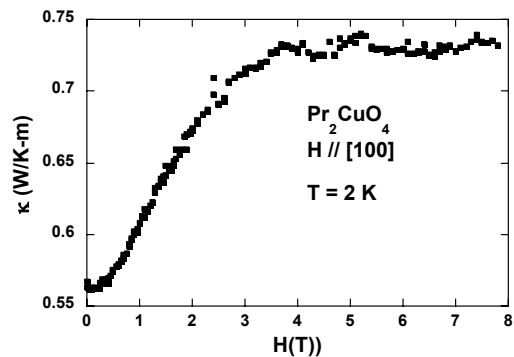

Figure 3. In-plane thermal conductivity of $Pr_2CuO_4$ at T = 2 K with H applied along the [100] direction.

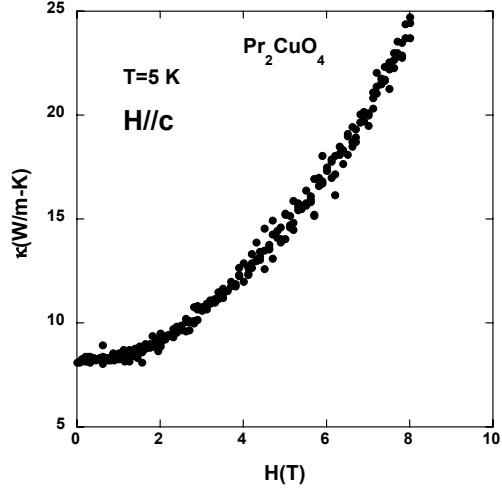

Figure 4. In-plane thermal conductivity of $Pr_2CuO_4$ at T = 5 K with the magnetic field applied along the *c* axis. There is no spin-flop transition for H // c. The increase in κ with field is likely associated with the coupling of magnetic excitons and phonons.

## DISCUSSION

All ordered magnetic compounds have a small gap in the magnon spectra at k=0 . This gap simply reflects the fact that the magnetic spins choose to point in a certain direction relative to the crystal structure. This gap (which is also the gap measured in a ferromagnetic or antiferromagnetic resonance experiment) reflects the *anisotropy* in the magnetic coupling strength, J, and can be much smaller than J. All three systems were chosen because the size of this gap could be closed with a magnetic field attainable with the PPMS ($H_c$ < 9 T) . Closure of this gap results in a spin-flop or spin-reorientation transition and a new magnetic ground state. It is evident from figures 1 and 2 that the spin-flop transition can have a significant effect on the measured thermal conductivity, hence the term "Thermal Spin Valve".

Empirically we have found that the increase in the thermal conductivity as a function of magnetic field (H > $H_c$) and temperature is proportional to:

$$p_m(T,H) = \frac{e^{-(\Delta - g\mu H)/k_BT}}{e^{-(\Delta - g\mu H)/k_BT} + e^{-(\Delta + g\mu H)/k_BT}} - \frac{1}{1 + e^{-2\Delta/k_BT}} \quad (1)$$

This equation is a phenomenological way of quantifying the fraction of the system in the new magnetic ground state. This equation can also be regarded as simply reproducing the correct temperature and magnetic field dependence of the measured effect for magnetic fields larger than $H_c$. Initially we interpreted the increase in kappa for $H > H_c$ as evidence for additional conduction of heat by magnons[9,10]. A detailed theoretical analysis of the change in the magnon spectra (and magnon density of states) in $K_2V_3O_8$, however, was recently completed by A. N. Sasha Chernyshev [16]. Using the model Hamiltonian developed by Lumsden et al. to explain the neutron scattering results, Chernyshev showed that the increase in kappa for $H > H_c$ is associated with a *decrease* in the magnon density of states near k=0. He was able to semi-quantitatively reproduce all of the magneto-thermal conductivity features for $K_2V_3O_8$ assuming that the heat is carried primarily by phonons. For $H > H_c$ there is a reduction in the amount of phonon-magnon scattering and hence an increase in the measured thermal conductivity. A similar argument may be valid for $Nd_2CuO_4$ and $Pr_2CuO_4$, although, as noted above, the spin excitations in these compounds are much more complicated because of two coupled magnetic systems (Cu and rare earth spins). In addition, because of the large value of the magnetic coupling in the cuprates, the acoustic magnon velocity is much larger than the phonon velocity. In the cuprates substantial heat conduction by magnons is more plausible and seems to have been unambiguously demonstrated for the insulating cuprate $Ca_9La_5Cu_{24}O_{41}$ [6].

**SUMMARY**

A conventional spin valve is a layered device in which the relative orientation of the magnetization in the layers controls the electrical conductivity. By analogy, a thermal spin valve is a layered material in which the orientation of the magnetic moments relative to the layers controls the thermal conductivity. For $K_2V_3O_8$ and $Nd_2CuO_4$ a spin-flop transition has a clear effect on the low temperature thermal conductivity (Figs 1, 2). A theory developed by Sasha Chernyshev [16] for $K_2V_3O_8$ strongly suggests that for this compound the heat is carried by phonons and the increase in the thermal conductivity for $H > H_c$ is due to a reduction in phonon-magnon scattering. The magneto-thermal conductivity data from $Pr_2CuO_4$ show a dramatic increase in $\kappa_{ab}$ for H//a *and for H//c*. Since there is no spin-flop for H //c, it is likely that *other* magnetic excitations involving the Pr ions, such as magnetic excitons, are responsible for the effects shown in Fig. 3 and 4. The data shown in Fig 4. brings into question the conclusions of Hill et al. [8] concerning the breakdown of Fermi liquid theory and the Wiedemann-Franz law in $Pr_{1.85}Ce_{0.15}CuO_4$.